\documentclass[conference]{IEEEtran}
\IEEEoverridecommandlockouts
\usepackage{cite}
\usepackage{multicol}
\usepackage{multirow}
\usepackage{array}
\usepackage{booktabs}
\usepackage{amsmath,amssymb,amsfonts}
\usepackage{algorithmic}
\usepackage{graphicx}
\usepackage{textcomp}
\usepackage{xcolor}
\usepackage{color,soul}
\usepackage{fancyhdr,lipsum}
\fancypagestyle{firstpage}{
  \fancyhf{}
  \fancyfoot[C]{978-0-7381-0508-6/21/\$31.00 ©2021 IEEE}
}

\graphicspath{{./images/}}
\def\BibTeX{{\rm B\kern-.05em{\sc i\kern-.025em b}\kern-.08em
    T\kern-.1667em\lower.7ex\hbox{E}\kern-.125emX}}
\newtheorem{remark}{Remark}

\begin{document}

\title{Deep Learning-Aided 5G  Channel Estimation\\
}


\author{\IEEEauthorblockN{An Le Ha$^{\dagger,\ast}$, Trinh Van Chien$^{\nu,\xi}$, Tien Hoa Nguyen$^\dagger$, Wan Choi$^\ast$, and Van Duc Nguyen$^{\dagger}$}
\IEEEauthorblockA{$^\dagger$School of Electronics and Telecommunications, Hanoi University of Science and Technology, Hanoi, Vietnam\\
$^\nu$School of Information and Communication Technology, Hanoi University of Science and Technology, Hanoi, Vietnam\\
$^\xi$Interdisciplinary Centre for Security, Reliability and Trust (SnT), University of Luxembourg, Luxembourg\\
$^{\ast}$Department of Electrical and Computer Engineering, Seoul National University, Seoul, Korea\\
Emails: an.lh150005@sis.hust.edu.vn,
vanchien.trinh@uni.lu, hoa.nguyentien@hust.edu.vn,\\ wanchoi@snu.ac.kr,
duc.nguyenvan1@hust.edu.vn
}
\thanks{This  paper  was  presented at  the 15th International Conference on Ubiquitous Information Management and Communication (IMCOM 2021).  ©2021  IEEE.  Personal  use  of  this  material  is permitted. Permission from IEEE must be obtained for all other uses, in any current  or  future  media,  including  reprinting/republishing  this  material  for advertising or promotional purposes, creating new collective works, for resale or redistribution to servers or lists, or reuse of any copyrighted component of this work in other works.}
}

\maketitle
\thispagestyle{firstpage}
\begin{abstract}
Deep learning has demonstrated the important roles in improving the system performance and reducing computational complexity for $5$G-and-beyond networks. In this paper, we propose a new channel estimation method with the assistance of deep learning in order to support the least squares estimation, which is a low-cost method but having relatively high channel estimation errors. This goal is achieved by utilizing a MIMO (multiple-input multiple-output) system with a multi-path channel profile used for simulations in the $5$G networks under the severity of Doppler effects. Numerical results demonstrate the superiority of the proposed deep learning-assisted channel estimation method over the other channel estimation methods in previous works in terms of mean square errors.
\end{abstract}


\section{Introduction}
The fifth-generation (5G) wireless communication has been developed to adapt to the exponential increases in wireless data traffic and reliability communications \cite{van2017massive}. The orthogonal frequency division multiplexing (OFDM) technique has been demonstrating its inevitable successes in the current networks, and  has continuously adopted in 5G systems to combat the frequency selective fading in multi-path propagation environments \cite{Bjornson2011a}. Consequently, this technique increases the spectrum efficiency compared with single-carrier techniques. Through the wireless multipath channels, the transmitted signals to a particular receiver is distorted by many detrimental effects such as multi-path propagation, local scattering, and mutual interference by sharing radio resources. Therefore, channel state information and its effects must be estimated and compensated at the receiver to recover the transmitted signals. Generally, pilot symbols known to both the transmitter and receiver are used for the channel estimation. In a $5$G system, the structure of the pilot symbols may be varied depending on different use cases \cite{dahlman20185g}. Among the conventional channel estimation methods, least squares (LS) estimation is a low computational complexity one since it requires no prior information of the statistical channel information. However, this estimation method yields relatively low performance in many application scenarios. Alternatively, the minimum mean square error (MMSE) estimation method has been introduced, which minimizes the channel estimation errors on average \cite{Kay1993a,Chien2018a}. The optimality of MMSE estimation is based on the assumption that the propagation channels are  modeled by a linear system and each channel response follows a circularly symmetric complex Gaussian distribution for which the channel estimates can be derived in the closed form \cite{mei2019machine,ha2016proposals}. Unfortunately, the MMSE estimation method has high computational complexity due to the requirements of channel statistic information, i.e., the mean and covariance matrices. In many environments, such statistical information is either difficult to obtain or quickly variant in a short time period \cite{nguyen2020performance,wu2014non}.

Machine learning has recently drawn much attention in various applications of wireless communications such as radio resource allocation, signal decoding, and channel estimation \cite{o2017introduction,van2019sum,van2019power, neumann2018learning}. Regarding the channel estimation, the authors in \cite{neumann2018learning} exploited the non-stationary channel conditions and the channel fading vectors are considered as conditionally Gaussian random vectors with random covariance matrices. The MMSE estimation form under those conditions may have an extremely high cost to obtain, and thus the authors used an estimation designed under a special channel condition for the machine learning aided estimation. In \cite{3}, the authors studied channel estimation in a wireless energy transfer system for which the downlink channel estimation is exploited to  harvest energy feedback information. A deep neural network model is used to predict better channel estimates than conventional estimations as LS or linear MMSE (LMMSE). These researches have numerically proved the compelling potentials of machine learning in channel estimation as long as sufficient training data set is provided. However, they only focused on the quasi-static propagation models such that channels are static and frequency flat in each coherence block.

In this paper, we propose two architectures of a deep neural network (DNN) model, which are applied for the channel estimation of a $5$G MIMO-OFDM system under frequency selective fading. The performance of the proposed deep learning-aided channel estimations is then evaluated by two different scenarios based on the receiver velocity. The channel parameters in each scenario are generated based on the tapped delay line type A model (TDL-A), which is reported by $3$GPP and of practical scenarios \cite{5}. The performance of the two DNN-aided channel estimations is compared with the traditional estimations, i.e., LS and Linear MMSE (LMMSE), in terms of mean square error (MSE) and bit error rate (BER) versus signal to noise ratio (SNR) criteria.\footnote{This paper uses LMMSE estimation as a benchmark for comparison because the channel estimates by MMSE estimation are nontrivial to obtain for the considered channel profile.} In particular, the proposed DNN structure will exploit a fully connected neural network to learn the features of actual channels with the channel estimates obtained by LS estimation as the input. In comparison to LS estimation, we would like to evaluate how much the system performance is improved by the assistance of a DNN. Furthermore, we would like to observe if the proposed estimation can beat the performance obtained by LMMSE.

The rest of this paper is organized as follows: Section~\ref{Sec:Syst} describes the 5G MIMO-OFDM system model. Section~\ref{Sec:III} presents the problems of the conventional channel estimation methods and proposes the DNN-aided methods to solve these problems. Meanwhile, Section~\ref{Sec:IV} shows the simulation results that evaluate the performance of the proposed methods and compare with the other benchmarks. Finally, the main conclusions of this paper are presented in Section~\ref{Sec:V}.

\section{5G NEW RADIO MIMO-OFDM SYSTEM} \label{Sec:Syst}
\begin{figure*}
    \centering
    \includegraphics[trim=8.1cm 7.2cm 20.8cm 5.5cm, clip=true, width=7.0in]{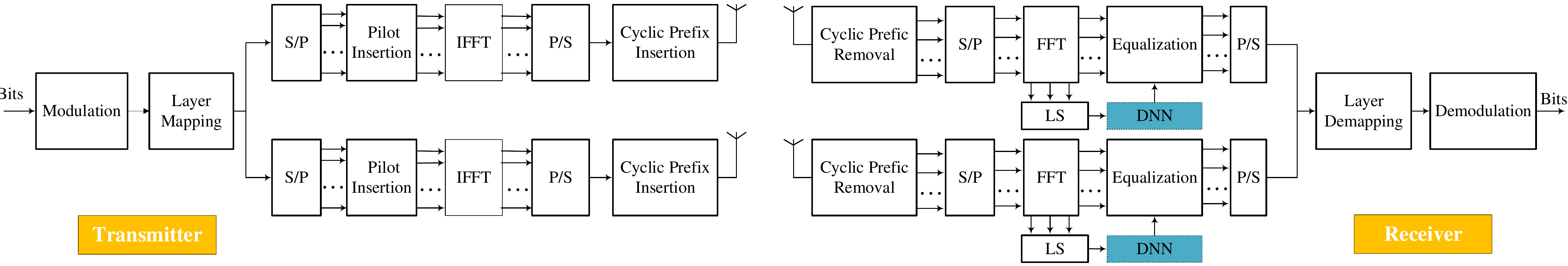}
    \caption{The considered MIMO-OFDM system model with the transmitter and receiver. The proposed DNN-aided module is in blue color. Notations: CP is Cyclic Prefix; S/P is Serial to Parallel;  P/S is Parallel to Serial; IFFT is Inverse Fast Fourier Transform; FFT is Fast Fourier Transform.}
    \label{Fig:SystemModel}
\end{figure*}

\begin{figure}
    \centering
    \includegraphics[trim=4cm 12.3cm 12.3cm 2cm, clip=true, width=3.5in]{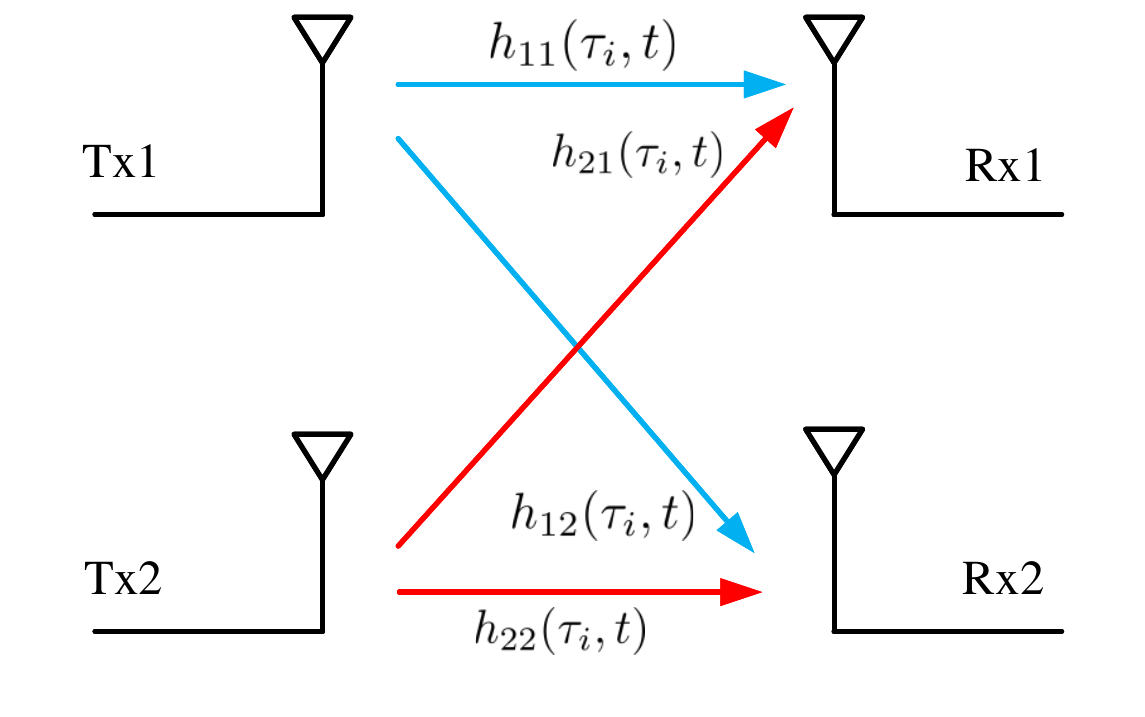}
    \centering
    \caption{The MIMO $2\times 2$ channel model where Tx$1$ and Tx$2$ are the transmit antenna indices, while Rx$1$ and Rx$2$ are the receive antenna indices.}
    \label{ChannelModel}
\end{figure}

In this paper, we consider a MIMO-OFDM system comprising a  transmitter sending signals to a receiver as illustrated in Fig.~\ref{Fig:SystemModel}. Both the transmitter and receiver are equipped with two antennas and therefore creating a $2\times 2$ MIMO channel model as in Fig.~\ref{ChannelModel}.

\subsection{Transmitter}
At the transmitter side, the binary data are mapped to the constellation points by utilizing the modulation block that exploits a modulation scheme such as quadrature amplitude modulation (QAM). We suppose that the system needs $T$ time slots to transmit the data and the QAM symbols at time slot~$t$,  $t = 1, \cdots, T,$  are combined to a data vector $\mathbf{x}(t) \in \mathbb{C}^N$ as
\begin{equation}
\mathbf{x}(t) = [x_1(t), x_2(t), \cdots, x_N(t) ],
\end{equation}
where $N$ is the total number of the modulation symbols. Then, the layer mapping block will separate vector $\mathbf{x}(t)$  into the two vectors $\mathbf{x}^{\mathrm{odd}}(t)$ and $\mathbf{x}^{\mathrm{even}}(t)$ corresponding to the two transmit antennas as follows:
\begin{align}
    &\mathbf{x}_1(t) = \mathbf{x}^{\mathrm{odd}}(t) = [x_1(t),x_3(t),\cdots,x_{N-1}(t)], \\
    &\mathbf{x}_2(t) = \mathbf{x}^{\mathrm{even}}(t) = [x_2(t),x_4(t),\cdots,x_N(t)].
\end{align}
The signals for each antenna are then converted from the serial to parallel one. At the pilot insertion block, the pilot symbols known by both the transmitter and receiver are inserted along with data subcarriers in every layer for the channel estimation purposes. Let us denote $\mathbf{x}_a (t)$ with $a=1,2,$ the signal vectors obtained after the pilot insertion block, then the IFFT (inverse fast Fourier transform) block is applied to $\mathbf{x}_a(t)$ such that the signals are transformed from the frequency domain into time domain (denoted by $\tilde{\mathbf{x}}_a(t)$) as
\begin{equation} 
\begin{split}
    \tilde{\mathbf{x}}_a(t) = \mathrm{IFFT} \{ \mathbf{x}_a(t)\}. 
\end{split}
\end{equation}
To avoid inter-symbol interference, a cyclic prefix of the length $N_G$ is inserted in each OFDM symbol by utilizing the CP (cyclic prefix) insertion block. So the transmitted signal including cyclic prefix, denoted by $\tilde{\mathbf{x}}_{ga}(t)$, is represented in time domain as follows:
\begin{equation}
    [\tilde{\mathbf{x}}_{ga}(t) ]_n = 
    \begin{cases}
    [\tilde{\mathbf{x}}_a(t)]_{n+N_{\mathrm{FFT}}} & n = -N_G,-N_G+1,\ldots,-1 \\
    [\tilde{\mathbf{x}}_a(t)]_n & n = 0,1,\ldots,N_{\mathrm{FFT}}-1,
    \end{cases}
\end{equation}
where $N_{\mathrm{FFT}}$ denotes the FFT size.
In more detail, the last $N_G$ samples of $\tilde{\mathbf{x}}_a(t)$ are used as cyclic prefix and inserted to the beginning of this symbol, resulting in the signal $\tilde{\mathbf{x}}_{ga}(t)$ with length of $N_{\mathrm{FFT}}+N_G$.
\subsection{5G Channel Model}
The 5G MIMO channel model is depicted in Fig.~\ref{ChannelModel} with the two transmit antennas and two receive antennas. This paper exploits the multipath fading channel model, which is time-variant and frequency selective. We denote the time-variant channel impulse response from the $a$-th transmit antenna to $b$-th receive antenna ($b=1,2$) is $h_{a,b}(\tau_i,t)$, where $\tau_i$ is the transmission delay at the $i$-th path. As reported in \cite{4}, the time-variant channel impulse response is modulated using the Monte-Carlo method as
\begin{equation} \label{eq:ChannelEq}
    h_{a,b}(\tau_i,t) = \frac{1}{\sqrt{M}}\sum_{i=1}^{L}\rho(i)\sum_{l=1}^{M}e^{j((2\pi f_{a,b,l,i}t+\theta_{a,b,l,i}}\delta(\tau - \tau_i),
\end{equation}
where $M$ is the number of harmonic functions. $L$ is the total number of paths for which $i=1,\ldots, L$.  The discrete Doppler frequency and Doppler phase are respectively defined as
\begin{align}
& f_{a,b,l,i} = f_{d,\max}\sin(2\pi u_{a,b,l,i}), \\
& \theta_{a,b,l,i} = 2\pi u_{a,b,l,i},
\end{align}
where $f_{d,\max}$ is the maximum Doppler frequency. The channel impulse response are simulated based on uniformly independent random variables $u_{a,b,l,i}$ in the range  $[0, 1]$. In \eqref{eq:ChannelEq}, $\rho(i)$ is the linear delay power at the $i$-th path. In particular, the TDL-A model defined by 3GPP standard \cite{5} for 5G channel model are exploited as reference power delay profile (PDP). Consequently, the transmitted signal after passing through the 5G multi-path channel is formulated as
\begin{equation} \label{eq_2}
    \tilde{\mathbf{y}}_{gb}(t) = \sum_{a \in \{1,2 \}} \mathbf{\tilde{h}}_{a,b}(\tau,t)\otimes \tilde{\mathbf{x}}_{ga}(t)+ \mathbf{\tilde{w}}_b(t),
\end{equation}
where $\mathbf{h}_{a,b}(\tau,t) = [h_{a,b}(\tau_1,t),\ldots,h_{a,b}(\tau_L,t)]$;  $\mathbf{\tilde{w}}_b(t)$ is additive noise vector, whose elements are independent and identically distributed random variables following a circularly symmetric complex Gaussian distribution with zero-mean and variance $\sigma_w^2$. $\otimes$ is the convolutional operator.

\subsection{Receiver}
At the receiver, the cyclic prefix is first removed out from the received signal $\tilde{\mathbf{y}}_{gb}(t)$ on each antenna by the cyclic prefix removal module to obtain the vector $\tilde{\mathbf{y}}_{b}(t)$ of the length $N_{\mathrm{FFT}}$. The signals are then split into parallel subcarriers and transformed into frequency-domain by the FFT block, which gives the frequency-domain signal $\mathbf{y}_b(t)$ as
\begin{equation} \label{eq:ReceivedSig}
    \mathbf{y}_b(t) = \mathrm{FFT} \{ \tilde{\mathbf{y}}_b(t) \}.
\end{equation}
The received pilot signal is exacted from frequency-domain signal for channel estimation purposes. Then, the processed signal $\mathbf{y}_b(t)$ is equalized and congregated into a serial sequence from all the receive antennas by the layer demapping block. The signal is further demodulated by the demodulation scheme that corresponds to what the transmitter has used. At this point, the output of the MIMO-OFDM system model is obtained as the final binary data sequence.

\section{Proposed DNN-Aided Channel Estimation} \label{Sec:III}
The coherent detection used in wireless communications needs knowledge of the propagation channels between the transmitter and the receiver, which are able to traditionally obtain by utilizing a channel estimation technique. This section presents the two widely-used channel estimation techniques, which motivates us to exploit a DNN architecture to mitigate the channel estimation errors.
\subsection{Motivation}
\begin{figure}[t]
    \centering
    \includegraphics[trim=0.0cm 0cm 0cm 0cm, clip=true, width=2.6in]{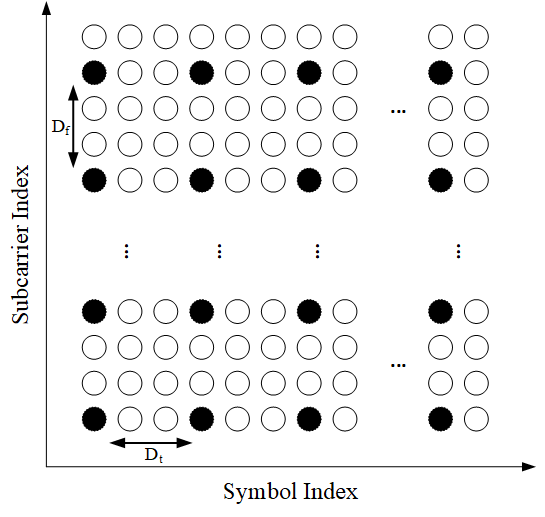}
    \caption{The pilot structure applied for the considered MIMO-OFDM system.}
    \label{Fig:PilotStructure}
\end{figure}

As long as there is no inter-carrier interference occurs, each subcarrier can be expressed as an independent channel, and therefore preserving the orthogonality among the subcarriers. The orthogonality allows each subcarrier component of the signal in \eqref{eq:ReceivedSig} to be expressed as the Hadamard product of the transmitted
signal and channel frequency response at the subcarrier \cite{book-cho} as
\begin{equation} \label{eq_3}
    \mathbf{y}_b(t) = \sum_{a \in \{1,2 \} }\mathbf{h}_{a,b}(t)\odot \mathbf{x}_a(t)+\mathbf{w}_a(t),
\end{equation}
where $\odot$ is the Hadamard product. $\mathbf{w}_b(t)$, $\mathbf{h}_{a,b}(t),$ and $\mathbf{x}_a(t)$ are the Fourier transforms of noise, channel, and signal respectively (or we are working in frequency domain). In a conventional estimation, the pilot symbols are supposed to know to both the transmitter and receiver are inserted along with data in frequency and time domain. In this paper, we apply the pilot structure of the 5G system defined in $3$GPP standard \cite{7-ts211}, which is shown in Fig.~\ref{Fig:PilotStructure}. The pilot symbols are uniformly spaced in the time domain, denoted by $D_t$ and in the frequency domain, denoted by $D_f$. The values of $D_t$ and $D_f$ depend on the different use cases of a 5G system, which are defined explicitly in, for example \cite{dahlman20185g}. 

Among all conventional channel estimation techniques, LS estimation is one of the most common. We denote $\hat{\mathbf{h}}_{\mathrm{LS}}$ by the channel estimate from the transmit antennas by this estimation technique. LS estimation gives the closed-form expression of the channel estimate as \cite{Kay1993a}:
\begin{equation}
    \hat{\mathbf{h}}_{\mathrm{LS}}(t) = (\mathbf{x}(t)^H\mathbf{x}(t))^{-1}\mathbf{x}^H(t)\mathbf{y}_b(t),
\end{equation}
where $(\cdot)^H$ denotes the Hermitian transpose, and
\begin{equation}
\mathbf{x}(t) = \big[\mathrm{diag}(\mathbf{x}_1(t)), \mathrm{diag}(\mathbf{x}_2(t)) \big]^T
\end{equation}
is the $N_{P} \times (2 N_{P})$ matrix denoting transmitted signal from the two transmit antennas; $N_P$ is the number of the pilot signals in an OFDM symbol; and $(\cdot)^T$ is the regular transpose. The channel estimate from each transmit antenna can be formulated as
\begin{multline}
\hat{\mathbf{h}}_{\mathrm{LSi}}(t) = \Big[ \big[ \hat{\mathbf{h}}_{\mathrm{LS}}(t) \big]_{(i-1)N_P}, \ldots, \big[\hat{\mathbf{h}}_{\mathrm{LS}}(t) \big]_{iN_P-1} \Big]^T,\\
i = 1,2.
\end{multline}
Then, the channel responses from all sub-carriers can be obtained by applying a linear interpolation method. Let us denote as $\hat{\mathbf{h}}_{\mathrm{LS}}(t)$. It should be noticed that LS estimation is a widely-used estimation because of its simplicity. However, this technique does not exploit the side information from noise and statistical channel properties in the estimation, and therefore high channel estimation errors might be obtained when applying LS estimation for complex propagation environments. 

To overcome the drawbacks, one can utilize LMMSE estimation, which minimizes the mean square error and having the channel estimation as \cite{book-cho}:
\begin{multline}
    \hat{\mathbf{h}}_{\mathrm{LMMSEi}}(t) = \mathbf{R}_{\mathbf{h}\hat{\mathbf{h}}_{\mathrm{LSi}}} \left(\mathbf{R}_{\mathbf{h}\mathbf{h}}+\frac{\sigma^2_w}{\sigma^2_x} \mathbf{I}_{N_P} \right)^{-1}\hat{\mathbf{h}}_{\mathrm{LSi}}(t), \\
    i=1,2,
\end{multline}
where $\hat{\mathbf{h}}_{\mathrm{LMMSEi}}(t)$ is the LMMSE estimated channel from the $i-$th transmit antenna, $\mathbf{R}_{\mathbf{h}\mathbf{h}} = \mathbb{E} \{\mathbf{h}\mathbf{h}^H\}$ is the autocorrelation matrix of channel response in frequency-domain with the size of $N_P \times N_P$ with $\mathbb{E}\{ \cdot \}$ being the expectation operator; $\mathbf{R}_{\mathbf{h}\hat{\mathbf{h}}_{\mathrm{LSi}}}= \mathbb{E}\{\mathbf{h}\hat{\mathbf{h}}_{\mathrm{LSi}}^H \}$ is the cross-correlation between the actual channel and the  channel estimate obtained by LS estimation with the size of $N_{\mathrm{FFT}} \times N_{P}$; $\sigma^2_x$ is the variance of the transmitted signals, respectively; $\mathbf{I}_{N_P}$ is the identity matrix of size $N_P \times N_P$. Since the impact of noise is taken into account by LMMSE estimation, which is able to improve the channel estimation accuracy. However, LMMSE estimation requires the prior knowledge of  channel statistical properties, thus the computational complexity is higher than LS estimation. Additionally, it may be difficult to obtain the exact distribution of channel impulse responses in general \cite{dnn-theoretical}, the performance of LMMSE estimation can not always be guaranteed. 
\begin{figure}[t]
    \centering
    \includegraphics[trim=3cm 4cm 3.0cm 3cm, clip=true, width=3.5in]{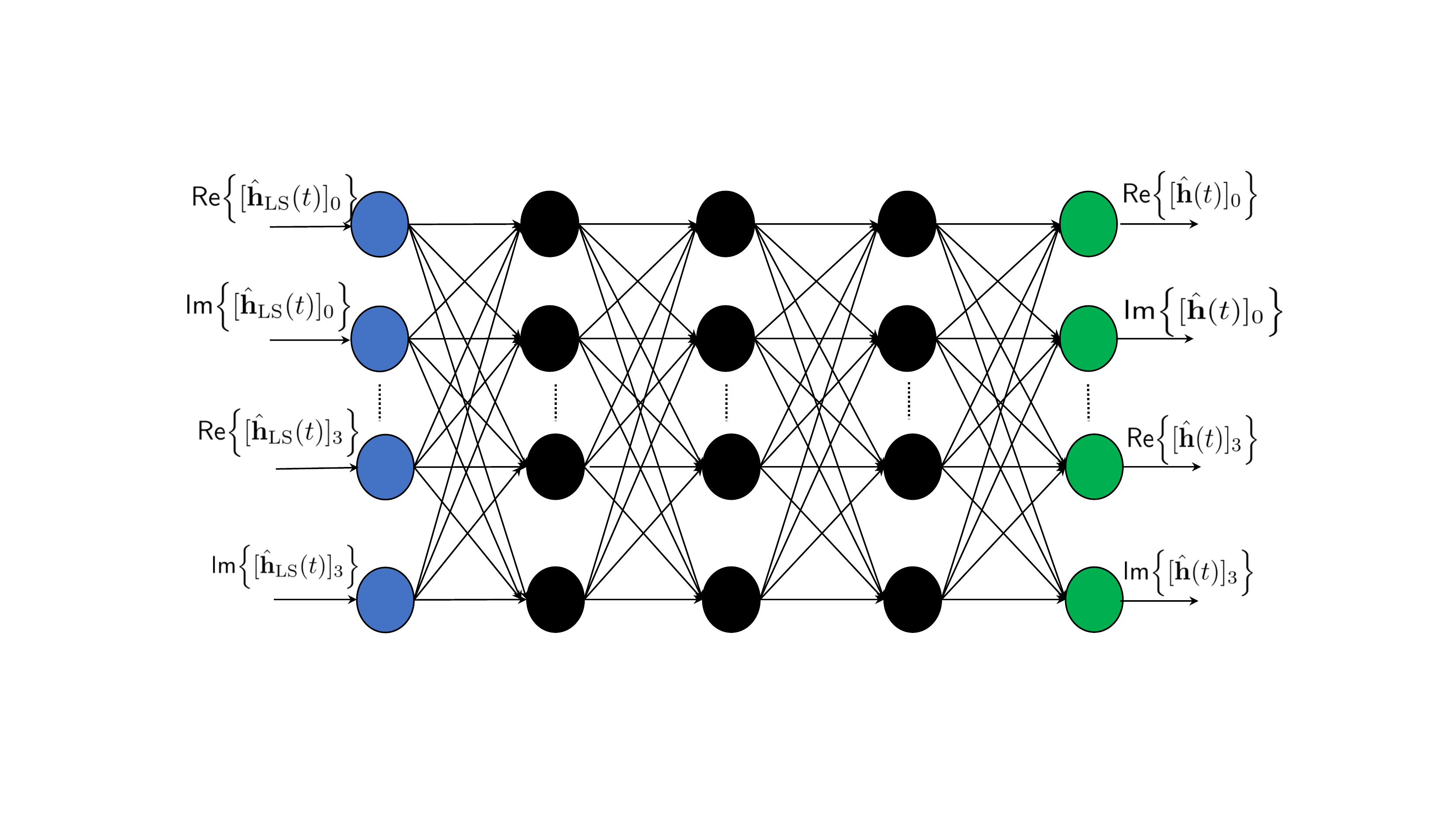}
    \centering
    \caption{The DNN structure used for channel estimation.}
    \label{fig_1}
\end{figure}
\subsection{DNN-Aided Channel Estimation}
To overcome the aforementioned drawbacks of LS and LMMSE estimations, we propose a DNN-aided estimation that minimizes the MSE between the channel estimate obtained by LS estimation and the actual channel.\footnote{According to the universal approximation theorem, there are other deep neural networks that give the similar or better performance than a fully-connected neural network with a limited data volume. However, the main theme of this paper is to point out the assistance of deep learning to channel estimation for 5G wireless communications. Therefore, the fully-connected DNN is selected due to its simplicity and low computational complexity. An optimized DNN structure is left for the future work.} The structure of the proposed DNN-aided estimation is depicted in Fig.~\ref{fig_1}. As shown in this figure, the proposed DNN structure is organized as layers including the input layer, hidden layers and output layer. Notice that a DNN may have many hidden layers. However, for the considered MIMO-OFDM system, the proposed DNN structure is designed with $3$ hidden layers which include multiple neurons. In particular, a neuron is a computational unit which performs the following calculation:
\begin{equation} \label{eq:Neuron}
    o = f(z) = f\left(\sum_{i=1}^{M}w_ix_i+b\right),
\end{equation}
where $M$ is the number of inputs to this neuron for which $x_i$ is the $i$-th input ($i=1, \ldots,M$); $w_i$ is the $i$-th weight corresponding to the $i$-th input;  $b$ is a bias and $o$ is the output of this neuron. In \eqref{eq:Neuron}, $f(.)$ is well-known as a activation function which is used to characterize the non-linearity of the data. In our proposed framework, we use the tanh function as the activation function, which is expressed as: \\
\begin{equation}
     f(z) = \frac{e^z-e^{-z}}{e^z+e^{-z}}.
\end{equation} 
To minimize the MSE, the DNN-aided estimation will learn the actual channel information given the channel estimates obtained by LS estimation as the input. In detail, we define a realization of the input for the training process as
\begin{multline} \label{eq:M}
 \mathcal{M}_{nt} = \left\{ \mathsf{Re} \Big\{ \big[ \hat{\mathbf{h}}_{\mathrm{LS}}^{n}(t) \big]_0 \Big\}, \mathsf{Im} \Big\{ \big[ \hat{\mathbf{h}}_{\mathrm{LS}}^{n}(t) \big]_0 \Big\}, \ldots,\right. \\
 \left.\mathsf{Re} \Big\{ \big[ \hat{\mathbf{h}}_{\mathrm{LS}}^{n}(t) \big]_{3} \Big\}, \mathsf{Im} \Big\{ \big[ \hat{\mathbf{h}}_{\mathrm{LS}}^{n}(t) \big]_{3} \Big\}\right\},
\end{multline}
 where the superscript $n$ denotes the $n$-th realization; $\mathsf{Re}\{ \cdot \}$ and $\mathsf{Im}\{ \cdot \}$ give the real and imaginary part of a complex number, respectively. The output of the neural network is 
 \begin{multline} \label{eq:O}
\mathcal{O}_{nt} = \left\{ \mathsf{Re} \Big\{ \big[ \hat{\mathbf{h}}^{n}(t) \big]_0 \Big\}, \mathsf{Im} \Big\{ \big[ \hat{\mathbf{h}}^{n}(t) \big]_0 \Big\}, \ldots,\right. \\
 \left.\mathsf{Re} \Big\{ \big[ \hat{\mathbf{h}}^{n}(t) \big]_{3} \Big\}, \mathsf{Im} \Big\{ \big[ \hat{\mathbf{h}}^{n}(t) \big]_{3} \Big\}\right\},
 \end{multline}
 where $\hat{\mathbf{h}}^n(t)$ is the output of the neural network at the $n$-th realization. In \eqref{eq:M} and \eqref{eq:O}, we separate the channel estimate into into the real and imaginary parts to tackle the complex numbers for the neural network. The learning process handles the one-by-one mapping:
 \begin{multline}
  \left(\mathsf{Re} \Big\{ \big[ \hat{\mathbf{h}}_{\mathrm{LS}}^{n}(t) \big]_s \Big\}, \mathsf{Im} \Big\{ \big[ \hat{\mathbf{h}}_{\mathrm{LS}}^{n}(t) \big]_s \Big\}\right) \rightarrow \\
   \left(\mathsf{Re} \Big\{ \big[ \hat{\mathbf{h}}^{n}(t) \big]_s \Big\}, \mathsf{Im} \Big\{ \big[ \hat{\mathbf{h}}^{n}(t) \big]_s \Big\}\right), s = 0, \ldots, 3,
 \end{multline}
As desired, the output of the neural network should be identical to the actual channels. Alternatively, the purpose of the DNN-aided estimation is to minimize the MSE between the prediction and  actual channels on average, thus the loss function utilized for the training phase is defined as
\begin{equation} \label{eq:LossFunction}
    \mathcal{L} \left( \mathcal{W}, \mathcal{B} \right) = \frac{1}{NT}\sum_{n=1}^{N} \sum_{t=1}^{T}\big\|\hat{\mathbf{h}}^{n}(t) - \mathbf{h}^{n}(t) \big \|_2^2,
\end{equation}
where $N$ is the number of realizations used for training, and $\mathbf{h}^{n}(t)$ is the actual channel corresponding to $\hat{\mathbf{h}}^{n}(t)$. $\mathcal{W}$ and $\mathcal{B}$ include all the weights and biases, respectively. From a set of initial values, the weights and biases are updated by minimizing the loss function \eqref{eq:LossFunction} with the forward and backward propagation \cite{van2019power}.
\begin{remark}
The loss function \eqref{eq:LossFunction} formulates a supervised learning. It is based on the fact that the actual channels are available in the training phase, which is obtained if the pilot power or coherence interval is sufficiently large. Consequently, the proposed learning-based approach possibly outperforms LS estimation. For future works, we can investigate the system performance under an unsupervised learning framework together with imperfect channel state information. 
\end{remark}

In order to train and test the proposed neural network, a set of data with $250880$ realizations  are gathered. We use  $70\%$ data for training, $15\%$ as the validation set, and $15\%$ data for testing.  In this paper, the two DNN models, labeled DNN-1 and DNN-2, are proposed for channel estimation with 5 layers comprising the input layer, $3$ hidden layers, and  output layer. As illustrated in Table.~\ref{Table1}, the number of neurons of each layer are $8$, $16$, $16$, $16$, $8$ for DNN-$1$, respectively. Meanwhile, those are  $8$, $32$, $32$, $32$, $8$ for DNN-$2$. Notice that the number neurons in input and output layers corresponds to the total number of real and image parts for 4 path channel, which is 8.\\
\begin{table}[t]
    \centering
    \caption{Architecture of DNN models for channel estimation}
        \begin{tabular}{*5l}
        \toprule
            \multirow{2}{5em}{Layer} &\multicolumn{2}{c}{DNN-1}&\multicolumn{2}{c}{DNN-2}  \\
            
             &Nodes & \(f(.)\) & Nodes & \(f(.)\) \\
        \midrule
             Input layer & 8 & - & 8 & - \\
             Hidden layer $1$ & 16 & tanh & 32 & tanh \\
             Hidden layer $2$ & 16 & tanh & 32 & tanh \\
             Hidden layer $3$ & 16 & tanh & 32 & tanh \\
             Output layer & 8 & - & 8 & - \\
        \bottomrule
        \end{tabular} \label{Table1}
\end{table}

\section{SIMULATION RESULTS} \label{Sec:IV}
To evaluate the performance of the DNN-aided estimation, the simulation has been carried out and the results are compared with the conventional LS estimation and LMMSE estimation by utilizing the bit error rate (BER) and mean square error (MSE) versus signal to noise ratio (SNR). The setup parameters of the considered MIMO-OFDM system are shown in Table~\ref{Table2}, while those parameters used for the DNN model are in Table~\ref{Table3}. In the simulations, we use the fading multi-path model channel with the TDL-A Power Delay Profile as aforementioned in Section~\ref{Sec:Syst}. For comparison, LS and LMMSE estimations are also included as benchmarks.

To investigate the performance of all the considered channel estimations exploiting in the MIMO-OFDM system through the 5G channel model, the two different scenarios corresponding to the velocity of mobiles are exploited:  In the first scenario, the receiver moves with a low speed such that the maximum Doppler frequency is $36$ Hz. The pilot symbols are inserted along with data in both frequency and time domain. Because the channel is slowly changed over time, the pilot spacing in the time domain is $D_t=4$ and in the frequency domain is $D_f = 2$; In the second scenario, the system serves high-speed mobility, which results in the maximum Doppler frequency of $200$ Hz. In this scenario, the setup $D_t = D_f = 2$ is to cope with a rapid change of the channels over time. 
\begin{table}[t!] 
    \centering
    \caption{Parameters for MIMO-OFDM system}
        \begin{tabular}{*2l}
        \toprule
            Parameters &Values \\
        \midrule
            MIMO & 2x2\\
            FFT size & 512\\
            Cyclic prefix & 64 \\
            Type of modulation & QPSK\\
            Channel PDP & TDL-A\\
            Maximum Doppler frequency & 36 Hz, 200 Hz\\
            Noise model &  Gaussian Noise \\
        \bottomrule
        \end{tabular} \label{Table2}
\end{table}
\begin{table}[t!]
    \centering
    \caption{Parameters for Deep neural network models}
        \begin{tabular}{*2l}
        \toprule
            Parameters &Values \\
        \midrule
            Training function & Levenberg-Marquardt\\
            Maximum number of epoches & $300$\\
            Mini-bath size & $8$ \\
            Training error & $10^{-5}$ \\
            Gradient descent accuracy & $10^{-7}$\\
            Learning rate & $0.01$\\
            Maximum validation failures & $6$\\
        \bottomrule
        \end{tabular} \label{Table3}
\end{table}
\begin{figure}[t]
    \centering
    \includegraphics[trim=0.2cm 0.0cm 0.2cm 0.2cm, clip=true, width=3.8in]{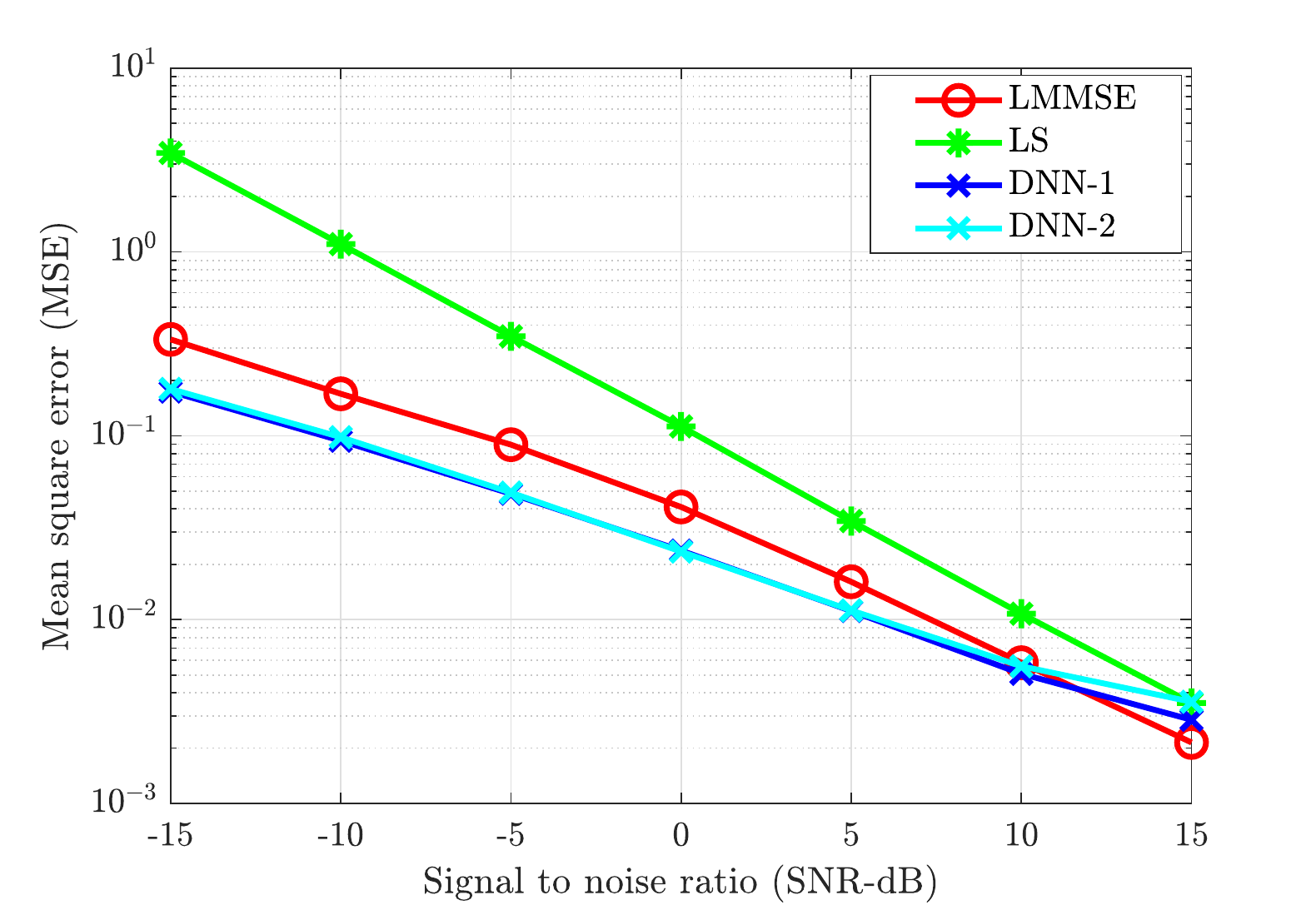}
    \centering
    \caption{The MSE of the channel estimate versus the SNR level for the first scenario $f_D = 36$ Hz.}
     \label{fig1}
\end{figure}
\begin{figure}[t]
    \centering
    \includegraphics[trim=0.2cm 0.0cm 0.2cm 0.2cm, clip=true, width=3.8in]{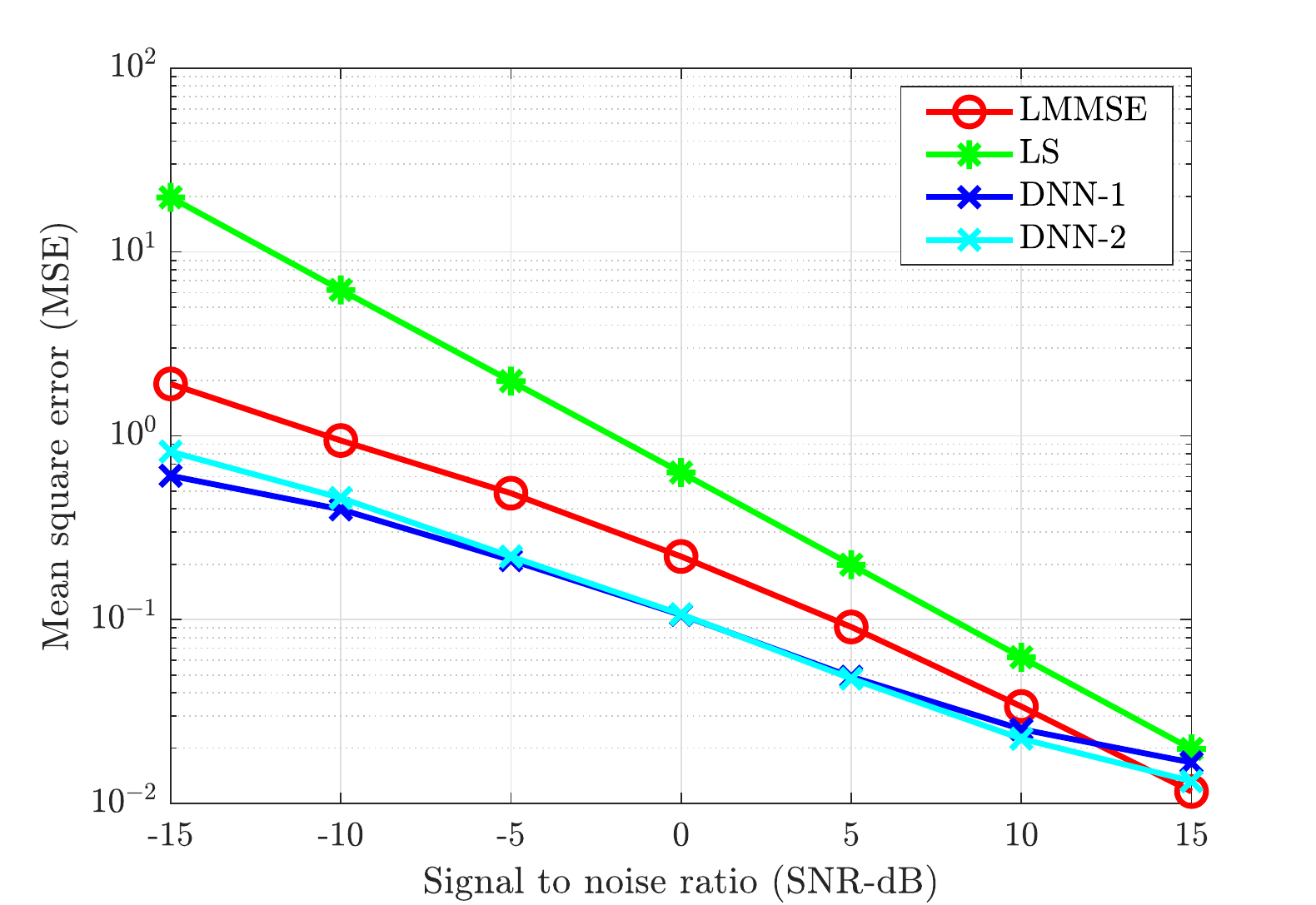}
    \centering
    \caption{The MSE of the channel estimate versus the SNR level for the second scenario $f_D = 200$ Hz.}
    \label{fig2}
\end{figure}
\begin{figure}
    \centering
    \includegraphics[trim=0.2cm 0.0cm 0.2cm 0.2cm, clip=true, width=3.8in]{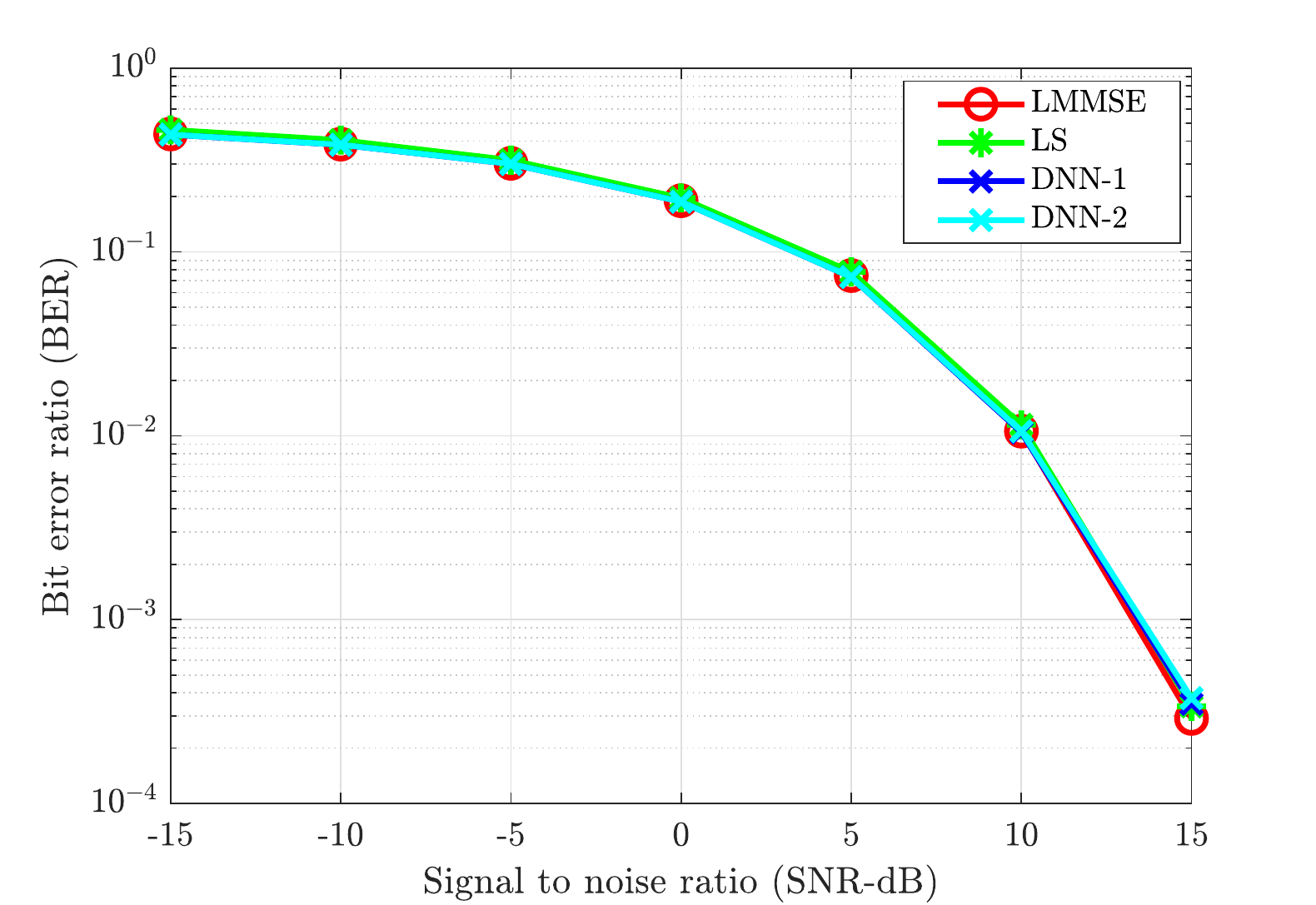}
    \centering
    \caption{The BER of the channel estimate versus the SNR level for the first scenario $f_D = 36$ Hz.}
    \label{fig3}
\end{figure}
\begin{figure}
    \centering
    \includegraphics[trim=0.2cm 0.0cm 0.2cm 0.2cm, clip=true, width=3.8in]{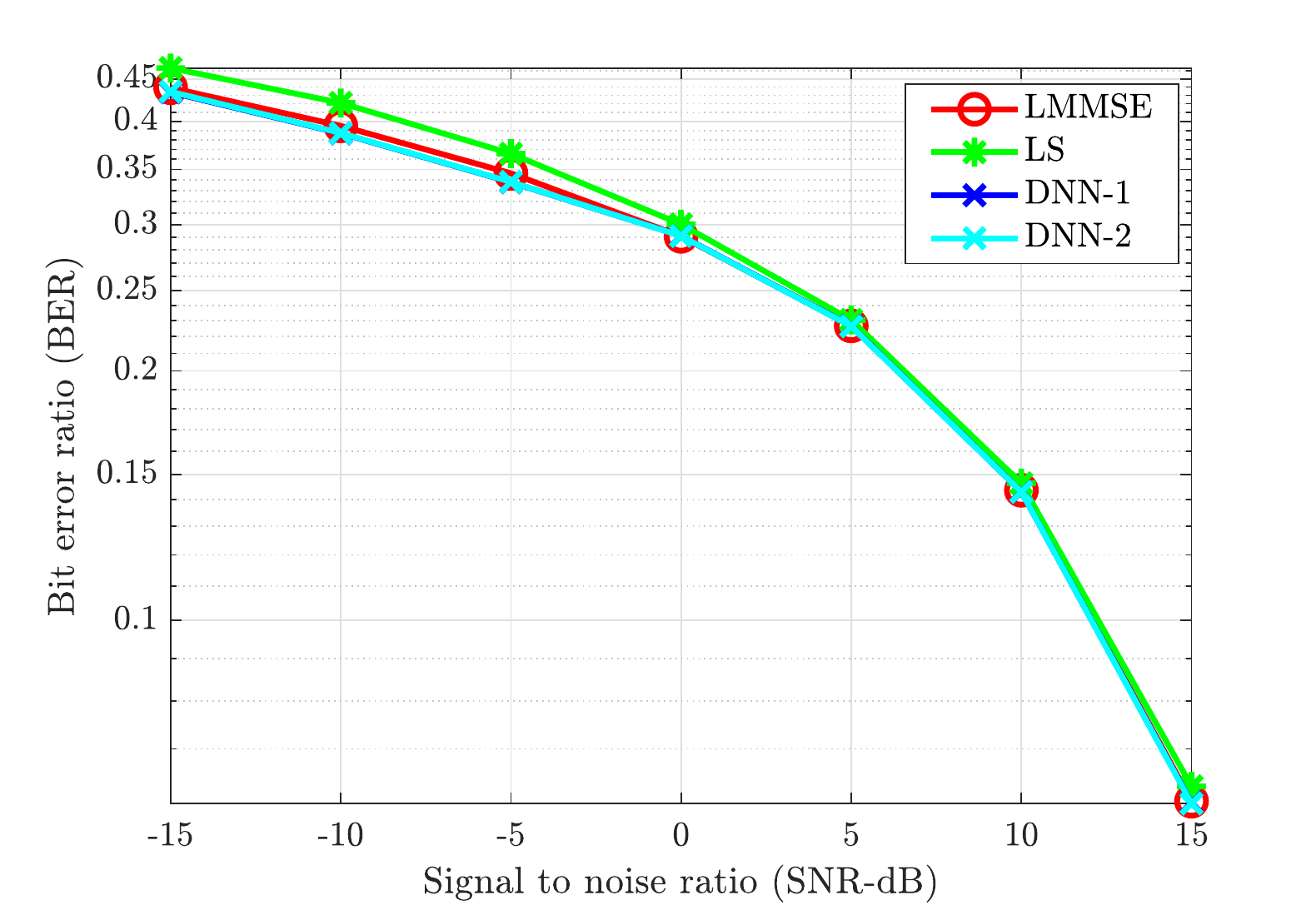}
    \centering
    \caption{The BER of the channel estimate versus the SNR level for the first scenario $f_D = 200$ Hz.}
    \label{fig4}
\end{figure}

Fig.~\ref{fig1} and \ref{fig2} show the MSE of different channel estimations utilizing the first and second scenarios, respectively. The QPSK (quadrature phase-shift keying modulation) is deployed to modulate the transmitted data in the simulation. As shown in Fig.~\ref{fig1} and \ref{fig2}, all the channel estimation methods provide MSE declining gradually as the SNR grows. In both  the scenarios, LS estimation yields the worst MSE performance since it does not take the statistical channel information into account when performing the channel estimation. On the contrary, LMMSE estimation exploits the mean and covariance matrices, which results in the better MSE performance than the LS counterpart. Our proposed deep learning methods yield the best MSE performance, especially at the low and mediate SNR levels. When the SNR increases above $13$ dB, the deep learning-based approaches give worse MSE than the performance of LMMSE estimation. This may be because the structure of the DNN models is still not optimal at high SNR levels and the hyper-parameters should be tuned more carefully. Although the DNN-$2$ model has more neurons in each hidden layer than the DNN-$1$ model, the results are only slightly different. This means that a complex DNN structure is not always along with  better accuracy. Although the pilot symbols are inserted more densely in time domain in the second scenario with the high speed of the receiver, the MSE of all four channel estimation methods is worse than those of the first scenario due to the severity of Doppler effects.

We provide the BER performance of the considered scenarios in Fig.~\ref{fig3} and \ref{fig4} with the different channel estimation methods, respectively. The  discrepancies across the channel estimation methods are not seen clearly in the BER performance. However, we still observe that LS estimation provides the worst performance among the four methods in both scenarios, while the BER performance of the remaining ones are almost the same to each other. Even though LS estimation performs worse than the others, the performance gap is relatively small. This can be explained by the fact that the loss function has been defined to minimize the channel estimation errors instead of the BER metric. Besides, Fig.~\ref{fig3} and \ref{fig4} also show the significant improvements of the BER when increasing the SNR level with combating the Doppler effects more effectively. For instance, at $f_D = 36$ Hz, the BER gets $10\times$ better if the SNR level increases from $5$~dB to $10$~dB.

\section{CONCLUSIONS} \label{Sec:V}
In this paper, the deep neural network with the two typical instances called DNN-1 and DNN-2 has been proposed to assist the channel estimation in a MIMO-OFDM system with the two different scenarios of fading multi-path channel models based on the TDL-A model defined in the 5G networks. The proposed DNN-based channel estimation methods are trained with the channel estimate from least squares estimation and the corresponding perfect channels. By utilizing the QPSK modulation scheme, the performance of the proposed estimations is compared with the conventional LS and LMMSE estimations in terms of channel estimation errors and bit error ratio as a function of the SNR levels. Due to learning the channel properties effectively, we observed the superior improvements of the proposed DNN-aided estimation in reducing channel estimation errors. The future work should focus on a design to reduce the bit error ratio as well. 
\section*{ACKNOWLEDGEMENT} \label{Sec:V}
This work was funded by the Vietnam’s Ministry of Education and Training (MOET) Project B2019-BKA-10.
\bibliographystyle{IEEEtran}
\bibliography{IEEEabrv,refs}
\end{document}